\documentclass[a4paper]{report}
\usepackage[utf8]{inputenc}
\usepackage[T1]{fontenc}
\usepackage{RJournal}
\usepackage{amsmath,amssymb,array}
\usepackage{booktabs}

\makeatletter
\DeclareRobustCommand*\textsubscript[1]{%
  \@textsubscript{\selectfont#1}}
\def\@textsubscript#1{%
  {\m@th\ensuremath{_{\mbox{\fontsize\sf@size\z@#1}}}}}
\makeatother

\begin{document}

\sectionhead{Contributed research article}
\volume{XX}
\volnumber{YY}
\year{20ZZ}
\month{AAAA}

\begin{article}
\title{\pkg{optimParallel}: an R Package Providing Parallel Versions of the Gradient-Based Optimization Methods of optim()}
\author{by Florian Gerber, Reinhard Furrer}

\maketitle

\abstract{
The R package \CRANpkg{optimParallel} \citep{optimParallel} provides a parallel version of the gradient-based optimization methods of \code{optim()}.
The main function of the package is \code{optimParallel()}, which has the same usage and output as \code{optim()}.
Using \code{optimParallel()} can significantly reduce optimization times.
We introduce the R~package and illustrate its implementation, which takes advantage of the lexical scoping mechanism of~R. 
}

\section{Introduction}
Many statistical tools involve optimization algorithms, which aim to find the minima or maxima of a function $fn: \mathbb{R}^p \rightarrow \mathbb{R}$, where $p\in\mathbb{N}$ denotes the number of parameters.
Depending on the specific application different optimization algorithms may be preferred; see the book by \cite{Nash:14}, the special issue of the Journal of Statistical Software \citep{jss:optim}, and the CRAN Task View \ctv{Optimization} for overviews of the optimization software available for~R. 
Widely used algorithms are the gradient-based optimization methods denoted by \mbox{\code{"L-BFGS-B"}}, \code{"BFGS"}, and \code{"CG"}, which are available through the general-purpose optimizer \code{optim()} of the R~package \pkg{stats}; see the help page of \code{optim()} for more information.
Those algorithms have proven to work well in numerous application. 
However, long optimization times of computationally intensive functions sometimes hinder their application; see \cite{Gerb:Mosi:Furr:16} for an example of such a function from our research in spatial statistics. 
For this reason we present a parallel \code{optim()} version in this article and explore its potential to reduce optimization times.

The gradient-based optimization algorithms of \code{optim()} rely on the evaluation of the gradient of $fn()$ denoted by $gr:\mathbb{R}^p \rightarrow \mathbb{R}^p$.
To illustrate the benefit of a parallel version of those algorithms we consider the optimization method \mbox{\code{"L-BFGS-B"}}, which alternately evaluates $fn()$ and~$gr()$.
Let $T_{fn}$ and $T_{gr}$ denote the evaluation times of $fn()$ and $gr()$, respectively. 
In the case where $gr()$ is specified by the user, one iteration of the algorithm evaluates $fn()$ and $gr()$ sequentially at the same parameter value.
Hence, the evaluation time of one iteration is $T_{fn}+T_{gr}$. 
In contrast, \code{optimParallel()} evaluates both functions in parallel using two processor cores, which reduces the evaluation time to little more than $\max\{T_{fn},T_{gr}\}$.
In the case where no gradient is provided, \code{optim()} calculates a numeric central difference approximation of $gr()$. 
For $p=1$ that approximation is defined as 
\begin{equation}
\widetilde{gr}(x)=(fn(x+\epsilon)-fn(x-\epsilon))\,/\,2\,\epsilon
\end{equation}
and hence requires two evaluations of $fn()$. 
Similarly, calculating $\widetilde{gr}()$ requires $2p$ evaluations of $fn()$ if $fn()$ has $p$ parameters. 
In total, \code{optim()} sequentially evaluates $fn()$ \,$1+2p$ times per iteration, resulting in an elapsed time of $(1+2p)T_{fn}$.
Given $1+2p$ available processor cores \code{optimParallel()} evaluates all calls of $fn()$ in parallel, which reduces the elapsed time to about $T_{fn}$ per iteration.

\section{\code{optimParallel()} by example}
The main function of the R~package \CRANpkg{optimParallel} is \code{optimParallel()}, which has the same usage and output as \code{optim()}, but evaluates $fn()$ and $gr()$ in parallel.
For illustration, consider $1000$ samples from a normal distribution with mean $\mu=5$ and standard deviation~$\sigma=2$. 
Then, we define the following negative log-likelihood and use \code{optim()} to estimate the parameters $\mu$ and~$\sigma$. 
\begin{example}
> x <- rnorm(n = 1000, mean = 5, sd = 2)
> negll <- function(par, x) -sum(dnorm(x = x, mean = par[1], sd = par[2], log = TRUE))
> o1 <- optim(par = c(1, 1), fn = negll, x = x, method = "L-BFGS-B", 
+             lower = c(-Inf, 0.0001))
> o1$par
[1] 5.017583 1.991459
\end{example}

Using \code{optimParallel()}, we can do the same in parallel.
The functions \code{makeCluster()}, \code{detectCores()}, and \code{setDefaultCluster()}, from the R~package \pkg{parallel} are used to set up a default cluster for \mbox{parallel~execution}. 
\begin{example}
> install.packages("optimParallel")
> library("optimParallel")
> cl <- makeCluster(detectCores()); setDefaultCluster(cl = cl)
> o2 <- optimParallel(par = c(1, 1), fn = negll, x = x, method = "L-BFGS-B",
+                     lower = c(-Inf, 0.0001))
> identical(o1, o2)
[1] TRUE
\end{example}

\begin{figure}[tb]
  \centering
  \includegraphics[width=.9\textwidth]{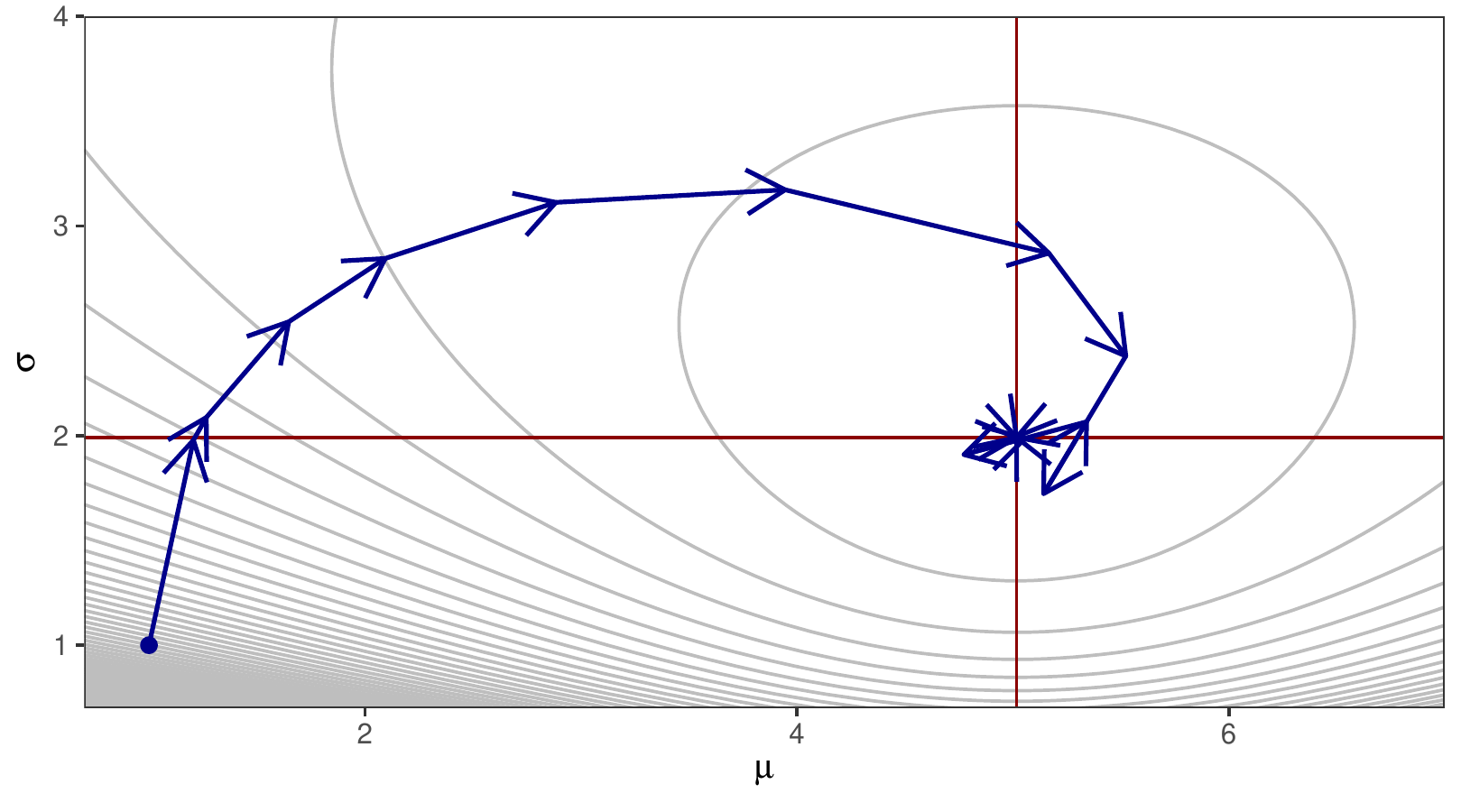}
  \caption{Visualization of the optimization path based on the log information obtained with \code{optimParallel(..., parallel = list(loginfo = TRUE))}.
The red lines mark the estimates $\hat \mu$ and $\hat \sigma$. 
}
  \label{fig:path}
\end{figure}

In addition to the arguments of \code{optim()}, \code{optimParallel()} has the argument \code{parallel}, which 
takes a list of arguments.
For example, we can set \code{loginfo = TRUE} to store the evaluated parameters and the corresponding gradients. 
\begin{example}
> o3 <- optimParallel(par = c(1, 1), fn = negll, x = x, method = "L-BFGS-B",
+                     lower = c(-Inf, 0.0001), parallel=list(loginfo = TRUE))
> print(o3$loginfo[1:3, ], digits = 3)
     iter par1 par2    fn   gr1    gr2
[1,]    1 1.00 1.00 10640 -3928 -18442
[2,]    2 1.21 1.98  3882  -951  -1801
[3,]    3 1.27 2.09  3655  -842  -1441
\end{example}
This can be used to visualize the optimization path as shown in Figure~\ref{fig:path}.

Another option is \code{forward}, which can be set to \code{TRUE} to enable the numeric forward difference approximation defined as $\widetilde{gr}(x)=(fn(x+\epsilon)-fn(x))/\epsilon\,$ for $p=1$ and $x$ sufficiently away from the boundaries.
Using that approximation, the return value of $fn(x)$ can be reused, and hence, the number of evaluations of $fn()$ is reduced to $1+p$ per iteration. 
This is helpful if the number of available cores is less that $1+2p$.

\section{Implementation}
\code{optimParallel()} is a wrapper to \code{optim()} and enables the parallel evaluation of all function calls involved in one iteration of the gradient-based optimization methods.
It is implemented in R and interfaces compiled C~code only through \code{optim()}. 
The reuse of the stable and well-tested C~code of \code{optim()} has the advantage that \code{optimParallel()} leads to the exact same results as \code{optim()}. 
To ensure that \code{optimParallel()} and \code{optim()} indeed return the same results \CRANpkg{optimParallel} contains systematic unit tests implemented with the R~package \CRANpkg{testthat} \citep{testthat,Hadl:11}.

The basic idea of the implementation is that calling \code{fn()} (or \code{gr()}) triggers the evaluation of both \code{fn()} and \code{gr()}.
Their return values are stored in a local environment. 
The next time \code{fn()} (or \code{gr()}) is called with the same parameters the results are read from the local environment without evaluating \code{fn()} and \code{gr()} again. 
The following R code illustrates how \code{optimParallel()} takes advantage of the lexical scoping mechanism of R to store the return values of \code{fn()} and \code{gr()}. 
\newpage
\begin{example}
> demo_generator <- function(fn, gr) {
+     par_last <- value <- grad <- NA
+     eval <- function(par) {
+         if(!identical(par, par_last)) {
+             message("--> evaluate fn() and gr()")
+             par_last <<- par
+             value <<- fn(par)  
+             grad <<- gr(par)   
+         } else 
+             message("--> read stored value")
+     }
+     f <- function(par) {
+         eval(par = par)
+         value
+     }
+     g <- function(par) {
+         eval(par = par)
+         grad
+     }
+     list(f = f, g = g)
+ }
> demo <- demo_generator(fn = sum, gr = prod)
\end{example}
Calling \code{demo\$f()} triggers the evaluation of both \code{fn()} and \code{gr()}.
\begin{example} 
> demo$f(1:5)
--> evaluate fn() and gr()
[1] 15
\end{example}
The subsequent call of \code{demo\$g()} with the same parameters returns the stored value \code{grad} without evaluating \code{gr()} again. 
\begin{example} 
> demo$g(1:5)
--> read stored value
[1] 120
\end{example}
A similar construct allows \code{optimParallel()} to evaluate $fn()$ and $gr()$ at the same occasion. 
It is then straightforward to parallelize the evaluations using the R~package \pkg{parallel}.

\section{Benchmark}
To illustrate the speed gains that can be achieved with \code{optimParallel()} we measure the elapsed times per iteration when optimizing the following test function and compare them to those of \code{optim()}.
\begin{example}
> fn <- function(par, sleep) {
+     Sys.sleep(sleep)
+     sum(par^2)
+ }
> gr <- function(par, sleep) {
+     Sys.sleep(sleep)
+     2*par
+ }
\end{example}
In both functions the argument \code{par} can be a numeric vector with one or more elements and the argument \code{sleep} controls the evaluation time of the functions. 
We measure the elapsed time per iteration for \code{method = "L-BFGS-B"} using all combinations of $p=1$, 
$2$, $3$, \code{sleep\,=} $0$, $0.05$, $0.2$, $0.4$, $0.6$, $0.8$, and $1$~seconds with and without analytic gradient \code{gr()}.
All measurements are taken on a computer with $12$ Intel Xeon E5-2640\,@\,2.50GHz processors. 
However, because of the experimental design maximum $7$~processors are used in parallel.
We repeat each measurement $5$ times using the R~package \CRANpkg{microbenchmark} \citep{microbenchmark}.
The complete R~script of the benchmark experiment is contained in \CRANpkg{optimParallel}.

\begin{figure}[tb]
  \centering
\hspace*{1cm}  \includegraphics[width=.6\textwidth]{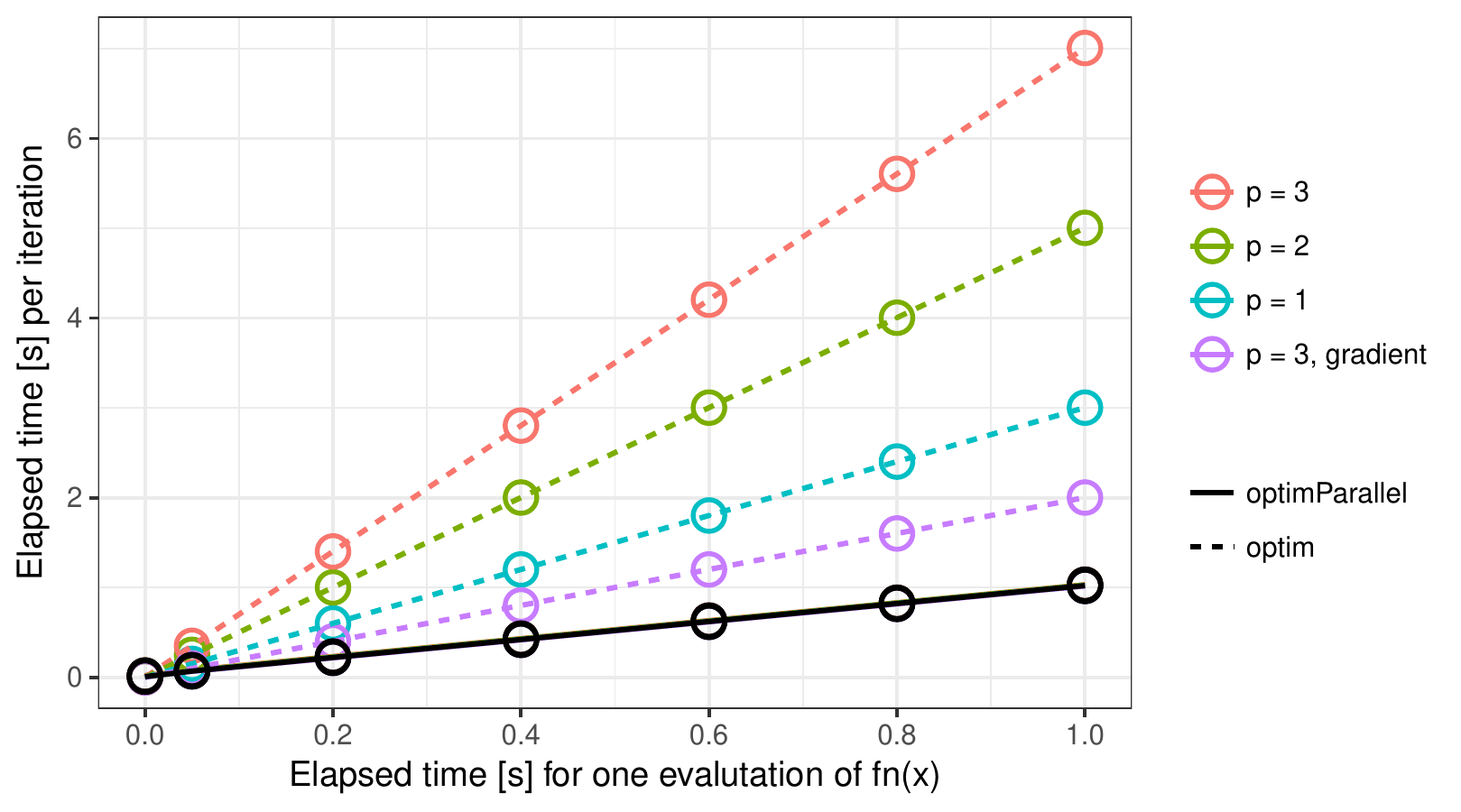}
  \caption{
Benchmark experiment comparing the \mbox{\code{"L-BFGS-B"}} method from \code{optimParallel()} and \code{optim()}. 
Plotted are the elapsed times per iteration ($y$-axis) and the evaluation time of $fn()$ ($x$-axis) for $p=1,\,2,$ and~$3$ using an approximate gradient and $p=3$ using an analytic gradient. 
The elapsed times of \code{optimParallel()} (solid line) are independent of~$p$ and the specification of an analytic gradient.
}
  \label{fig:bench}
\end{figure}

The results of the benchmark experiment are summarized in Figure~\ref{fig:bench}.
They show that for \code{optimParallel()} the elapsed time per iteration is only marginally larger than $T_{fn}$ (black dots in Figure~\ref{fig:bench}). 
Conversely, the elapsed time for \code{optim()} is $T_{fn}+T_{gr}$ if a gradient function is specified (violet dots) and
$(1+2p)T_{fn}$ if no gradient function is specified.
Moreover, \code{optimParallel()} adds a small overhead, and hence, it is only faster than \code{optim()} if $T_{fn}$ is larger than $0.05$ seconds.

\section{Summary}
The R~package \CRANpkg{optimParallel} provides parallel versions of the gradient-based optimization methods \mbox{\code{"L-BFGS-B"}}, \code{"BFGS"}, and \code{"CG"} of \code{optim()}.
After a brief theoretical illustration of the possible speed improvement based on parallel processing, we illustrate \code{optimParallel()} by examples.
The examples demonstrate that one can replace \code{optim()} by \code{optimParallel()} to execute the optimization in parallel and illustrate additional features like capturing log information and the forward gradient approximation.
Moreover, we briefly sketch the basic idea of the implementation, which is based on the lexical scoping mechanism of~R.
Finally, a benchmark experiment shows that using \code{optimParallel()} reduces the elapsed time to optimize computationally demanding functions significantly. 
For functions with evaluation times of more than $0.05$ seconds we measured speed gains of about factor~$2$ in the case where an analytic gradient was specified and about factor~$1+2p$ otherwise ($p$ is the number of~parameters).

\bibliography{optimParallel_arXiv}

\address{Dr.\ Florian Gerber\\
  Department of Mathematics, University of Zurich, Switzerland\\
  \email{florian.gerber@math.uzh.ch}, \url{https://orcid.org/0000-0001-8545-5263}
  }

\address{Prof.\ Dr.\ Reinhard Furrer\\
  Department of Mathematics \& Department of Computational Science, University of Zurich, Switzerland\\
  \email{reinhard.furrer@math.uzh.ch}, \url{https://orcid.org/0000-0002-6319-2332}
  }

\end{article}

\end{document}